\documentclass[%
reprint,
superscriptaddress,
 amsmath,amssymb,
 prl,
]{revtex4-2}
\usepackage[hidelinks]{hyperref}
\usepackage{graphicx}
\usepackage{dcolumn}
\usepackage{bm}

\begin{document}

\preprint{APS/123-QED}

\title{Large Hall angle of vortex motion in high-$T_c$ cuprate superconductors revealed by microwave flux-flow Hall effect}

\author{R. Ogawa}
\affiliation{
Department of Basic Science, the University of Tokyo, Meguro, Tokyo 153-8902, Japan
}
\author{F. Nabeshima}
\affiliation{
Department of Basic Science, the University of Tokyo, Meguro, Tokyo 153-8902, Japan
}
\author{T. Nishizaki}
\affiliation{
Department of Electrical Engineering, Kyushu Sangyo University, Fukuoka 813-8503, Japan
}
\author{A. Maeda}
\affiliation{
Department of Basic Science, the University of Tokyo, Meguro, Tokyo 153-8902, Japan
}
\date{\today}

\begin{abstract}
We investigated the nature of the quasi-particle state in the vortex core by means of the flux-flow Hall effect measurements at 15.8 GHz. 
We measured the flux-flow Hall effect in cuprate superconductors, Bi$_{2}$Sr$_{2}$CaCu$_{2}$O$_{y}$ and YBa$_{2}$Cu$_{3}$O$_{y}$ single crystals, whose equilibrium $B$-$T$ phase diagrams were different. 
As a result, we found that the Hall angle is independent of the magnetic field, and reaches an order of unity at low temperatures in BSCCO.  However, in YBCO, the angle increases with increasing magnetic field even at low temperatures.
We understood that this difference in the magnetic field dependence of the Hall angle is due to the difference in the influence of the pinning, which originated from the difference in the vortex state (liquid  vs. solid) between the two materials.
However, as a common feature, both materials showed a large tangent of the Hall angle at low temperatures, which was larger by an order of magnitude than those obtained in the effective viscous drag coefficient measurements. 
We discussed the origin of the discrepancy both  in terms of the possible nonlinearity of the viscous drag force and possible hidden dissipation mechanisms.
The unexpectedly large Hall angle of the vortex motion in cuprates revealed in our flux-flow Hall effect study poses a serious question on the fundamental understanding of the motion of the quantized vortex in superconductors, and it deserves further investigation.  
\end{abstract}

\maketitle

In type II superconductors under magnetic field, the magnetic field penetrates as a quantized flux, which is  accompanied by the surrounding rotating currents. Thus, it is called as a magnetic vortex.
In the central part of the vortex, called the vortex core, quasi-particles (QPs) are confined.
The confined QPs form quantized energy levels~\cite{Caroli1964}, which are expressed as
$
E_n = \hbar\omega_0 \left(n+1/2\right), 
$
$
\hbar\omega_0 \equiv \Delta^2/E_F,
$
where $n$ is an integer, and $\Delta$ and $E_F$ are the superconducting gap and the Fermi energy, respectively.
These quantized levels have a finite width, $\delta E\equiv \hbar/\tau$, where $\tau$ is the scattering time of the QPs in the vortex core.
The nature of the core depends on the ratio of the energy spacing $\Delta E\equiv \hbar\omega_0$ to its width $\delta E$, which is expressed as
$
r\equiv \Delta E/\delta E = \omega_0\tau.
$
When $r\ll$ 1 (dirty core), the core is almost like a normal metal.
In contrast, when $r\gg$ 1 (clean core), the quantized nature of the QPs becomes prominent.
In between these values, the core is called moderately clean.
The core of almost all conventional superconductors is dirty, because $\Delta E$ is extremely small (typically 0.01 K). 
On the other hand, the quantized nature are expected to be prominent in high-$T_c$ cuprate superconductors, since the energy gap is large ($\sim0.01$ eV) and the Fermi energy is small ($\sim0.1$ eV), leading to a large $\Delta E\sim10$ K.
Indeed, in the STS spectra of YBa$_{2}$Cu$_{3}$O$_{y}$~\cite{Maggio-Aprile1995} and Bi$_{2}$Sr$_{2}$CaCu$_{2}$O$_{y}$~\cite{Pan2000}, peaks in the dI/dV spectra were observed, which implies the  existence of the quantized levels.
However, this interpretation is somewhat controversial even at present (Ref.~\cite{Valles1991,Edwards1992,Miller1993,Nantoh1994,Maggio-Aprile1995,Shibata2003a,Shibata2010,Bruer2016,Berthod2017,Renner1998,Hoogenboom2000,Pan2000,Matsuba2003,Levy2005,Matsuba2007,Yoshizawa2013} and those cited therein).

Another approach for investigating the electronic structure in the core is to investigate the vortex motion to an external driving current, the flux flow~\cite{Bardeen1965,Nozieres1966,Blatter1994,Kopnin1976,Golosovsky1994}. 
In the conventional understanding, the equation of motion of a quantized vortex (for the magnetic field in the $z$ direction) under the driving current density $\bm{J}$, is expressed as
\begin{equation}
\kappa_p\bm{u}+\eta\dot{\bm{u}}+\alpha_{H} \dot{\bm{u}}\times\hat{z}=\Phi_0\bm{J}\times \hat{z}, \label{eqm}
\end{equation}
where $\bm{u}=(x, y)$ is the displacement of the vortex, $\kappa_p$ is a pinning constant, $\eta$ and $\alpha_H$ are the viscous drag coefficients in the longitudinal and transverse directions, respectively, $\hat{z}$ is the unit vector in the magnetic field direction, and $\Phi_0\equiv h/2e$ is the flux quantum. 
If the pinning can be neglected, the direction of the motion can be expressed in terms the Hall angle $\theta$ as 
\begin{equation}
\tan\theta=\frac{\left|\dot{x}\right|}{\left|\dot{y}\right|}=\frac{\alpha_{H}}{\eta}=\omega_0\tau
\end{equation}
(see supplemental material for details).
Thus, if we measure the flux flow in the longitudinal direction and transverse direction, namely, flux-flow Hall effect (FFHE), we will find the parameter $r=\omega_0\tau$.
Alternatively, an effective viscous drag coefficient,
\begin{equation}
\eta_{eff}=\eta+\frac{\alpha_{H}^2}{\eta}=\pi\hbar n\omega_0\tau,
\end{equation}
also provides information on $r=\omega_0\tau$. 
Indeed, $\eta_{eff}$ can be obtained by the surface impedance measurement by using a cylindrical cavity~\cite{Klein1993,Donovan1993,Dressel1993}.
It should be noted that the flux-flow resistivity $\rho_{eff}$ is expressed as $\rho_{eff}=\Phi_0 B/\eta_{eff}$.
From an experimental point of view, the latter is much easier.
Thus, there have been many studies of the flux flow of high-$T_c$ superconductors along this direction~\cite{Tsuchiya2001,Shibata2003,Hanaguri1999,Maeda2007a,Maeda2007}.
All of these show that the parameter $r$ is 0.1 $\sim$ 0.3, which means that the core of high-$T_c$ cuprate is moderately clean.
This is in contrast to the result expected from the STS data~\cite{Maggio-Aprile1995,Pan2000}.
In particular, in a series of YBCO crystals where the cleanness of the sample was changed systematically, 
the QP scattering time in the core was found to always be shorter than those outside the core~\cite{Maeda2007}.
Since this trend seems very universal, this suggests the existence of an universal mechanism of the extra dissipation for the vortex motion.
Although some mechanisms such as the Andreev reflection~\cite{Hofmann1998,Eschrig1999} or the spectral flow~\cite{Volovik1993,Hayashi1998} might contribute to this extra dissipation, there is no established understanding or any compelling explanations on what these results implicate.
In addition, a new dissipation mechanism where the inelastic relaxation contributes to conductivities in addition to the conventional Bardeen-Stephen conductivity has recently been proposed~\cite{smith2020}.
Therefore, the experimental investigation in a different manner is becoming important.

Bardeen and Stephen discussed FFHE of dirty superconductors, based on the scenario that Lorentz force acts on the vortex~\cite{Bardeen1965}.
In this model, the vortex core was the normal conductive core.
Thus, $\tan\theta$ is proportional to the magnetic field.
Another early phenomenological theory was given by Nozi\`{e}res and Vinen, where they consider the vortex core in the superclean limit~\cite{Nozieres1966}.
FFHE was referred hydrodynamically, and they presume that the Magnus force was the main force acting on the core.
They found that $\tan\theta$ was independent of the magnetic field.
After these early studies, microscopic theories were presented.  
In particular, calculations based on the Green function approach were perfomed ~\cite{Kopnin1976a,Kopnin1997}.
In these theories, $\tan\theta$ was independent of the magnetic field in the superclean regime.
In contrast in the moderately clean case, FFHE was represented as the sum of a small term, which was proportional to the magnetic field and another term, which was independent of the magnetic field.
In addition, many other theoretical calculations have been presented based on the quasi-classical equations~\cite{Larkin1995,Kopnin1995,Arahata2014,Kohno2016}.
We note that many of these theories are for temperatures that are sufficiently lower than the superconducting transition temperature $T_{c}$.

However, no work has experimentally investigated the FFHE study, except in the very vicinity of $T_{c}$~\cite{Hagen1990,Nagaoka1998}.
To discuss $r=\omega_0\tau$, we need to understand FFHE in the superconducting state.
In addition, it is necessary to eliminate the effect of pinning~\cite{Gittleman1968}. One way to eliminate it is the measurement by using DC pulses. However, it could cause Joule heating, which prevents accurate measurements.
Another way to measure FFHE by using high frequencies(typically microwaves) and analyze it using a model that includes the pinning effect~\cite{Coffey1991}, as was the case for the flux-flow measurement with cylindrical cavity.
In a previous paper, we developed a novel method to measure the microwave Hall effect in materials in the skin depth region, which enable us to measure FFHE~\cite{Ogawa2021}.

In this study, we investigate the microwave FFHE in cuprate superconductors, by using the recently developed method, and attempt to discuss the above mentioned mystery.
We measured two different superconductors; Bi$_{2}$Sr$_{2}$CaCu$_{2}$O$_{y}$ and YBa$_{2}$Cu$_{3}$O$_{y}$.
The former has the very strong anisotropy and exhibits vortex-liquid behavior in the wide region of the $B$-$T$ phase diagram, whereas the latter exhibits vortex-solid behavior in the major part of the $B$-$T$ diagram.
In fact, both materials exhibited very contrastive behavior for the magnetic field dependence of the Hall angle.
However, more importantly, both materials showed a large $r=\omega_0\tau$ at low temperatures, different from the effective viscous drag coefficient measurements.
Although we try to interpret the results in terms of the nonlinearity in the viscous drag coefficient or novel dissipation mechanisms not well understood yet,  results poses a serious question on the fundamental understanding of the motion of the quantized vortex in superconductors, and it deserves further investigation.  

\begin{figure}[htbp]
\centering
 	\includegraphics[bb=0 0 2156 1328, keepaspectratio,width=85mm]{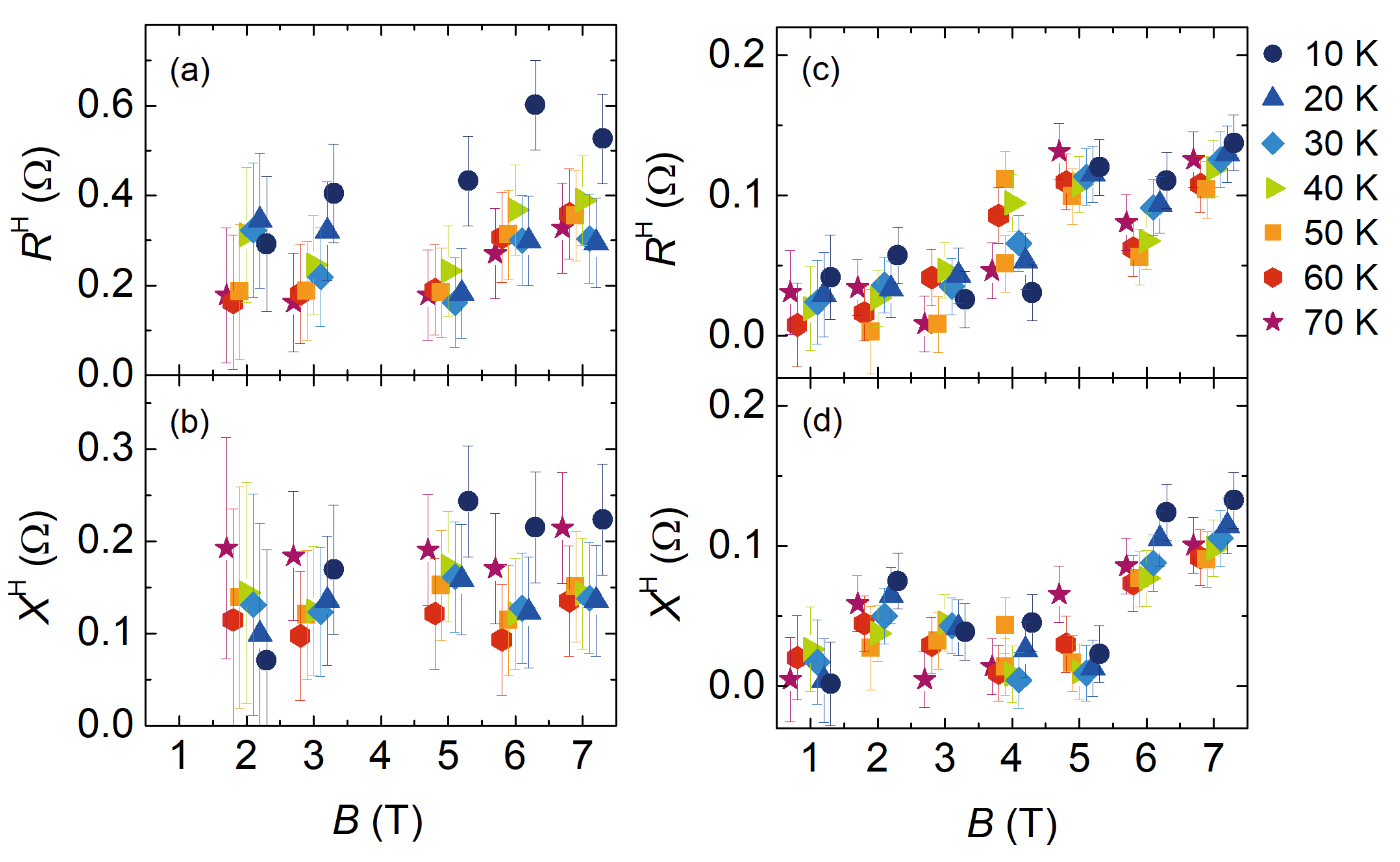}
 	\caption{
 	(a) Transverse surface resistance and (b) transverse surface reactance of BSCCO as a function of the magnetic field. 
 	(c) Transverse surface resistance and (d) transverse surface reactance of YBCO as a function of the  magnetic field. The signs of  $R^H$ and $X^H$ are set to be positive.
 	}
 	\label{cro}
\end{figure}

The tangent of the flux-flow Hall angle can be expressed as
$|\tan\theta|=|\dot{x}/\dot{y}|$.
Faraday's law ($\bm{E}=\bm{v}\times\bm{B}$) yields
$|\dot{x}/\dot{y}|=|E_y/E_x|$, 
where $\bm{E}$ denotes the electric field vector.
Since $\bm{E}=\tilde{\sigma}^{-1}\bm{J}\propto\tilde{Z}\bm{J}$, the tangent of the flux-flow Hall angle is given by
\begin{equation}\label{method_tan}
|\tan\theta|=\left|\frac{Z^H}{Z^L}\right|,
\end{equation}
where $\tilde{\sigma}$ denote the conductivity tensor; $Z^L$ and $Z^H$ are the longitudinal and transverse components of the surface impedance tensor $\tilde{Z}$, respectively.
Thus, we can obtain the Hall angle from the measurement of the longitudinal and off-diagonal component of the surface impedance tensor.
The off-diagonal component, $Z^H\equiv R^H-i X^H$ is obtained as follows, by using the cross-shaped bimodal cavity~\cite{Ogawa2021}.
The changes in the resonance characteristics of the bimodal cavity are represented as 
\begin{equation}\label{method_crossQ}
\Delta\left(\frac{1}{2Q_H}\right)=G^LR^L+G^H|X^H|,
\end{equation}
\begin{equation}\label{method_crossf}
\Delta \left(\frac{f_H}{f_{H0}}\right)\equiv -\frac{f_{H}-f_{H0}}{f_{H0}}=G^LX^L-G^H|R^H|,
\end{equation}
where $R^L$ and $X^L$ are the longitudinal components of the surface resistance and the surface reactance tensors, which are obtained in the ordinary cylindrical cavity measurement,  $R^H$ and $X^H$ are the off-diagonal components of the surface resistance and the surface reactance tensors, $G^L$ and $G^H$ are the geometrical constants, which depend on the shape of the bimodal cavity and sample, respectively, $\Delta$ represents the difference between the data of the same sample in the superconducting state and those in the normal state, which is understood very well.
Therefore, together with the cylindrical cavity measurement, whose detail is described in supplemental material, we can obtain flux-flow Hall angle.


we performed the measurement using the cross-shaped bimodal cavity, operating in the two orthogonal TE$_{011}$ and TE$_{101}$ modes at 15.8 GHz, which had the quality factor $Q\sim3\times 10^{3}$.
\ First, the microwave response was measured in the zero magnetic field.  Then, along with the surface impedance data obtained in the cylindrical cavity, the geometric factors were obtained as $G^L(=G^H)=3.5\times 10^{-6}$ $\Omega^{-1}$ for BSCCO and  $G^L(=G^H)=1.3\times 10^{-6}$ $\Omega^{-1}$ for YBCO. 
Since the geometrical factors were obtained, we could evaluate $R^H$ and $X^H$ from Eqs.~(\ref{method_crossQ}) and (\ref{method_crossf}) in the bimodal cavity measurement under finite magnetic fields.
Figs.~\ref{cro}(a) and (b) show the $R^H$ and $X^H$ of BSCCO from 10 K to 70 K as a function of the magnetic field. 
Because the sign of  $R^H$ and $X^H$ are not uniquely determined in the cross-shaped cavity measurement, we set these signs to be positive.
Both $R^H$ and $X^H$ do not show remarkable dependence on the magnetic field, except at low temperatures, and they increase with deceasing temperature.
Figs.~\ref{cro}(c) and (d) show the $R^H$ and $X^H$ of YBCO as a function of the magnetic field. 
Although the uncertainty was somewhat larger than that in the BSCCO case due to small signals, $R^H$ and $X^H$ of YBCO had essentially the same temperature dependence as those in the BSCCO, and they show dependence on the magnetic field.

\begin{figure}[htbp]
\centering
 	\mbox{\includegraphics[bb=100 0 700 600, keepaspectratio,width=85mm]{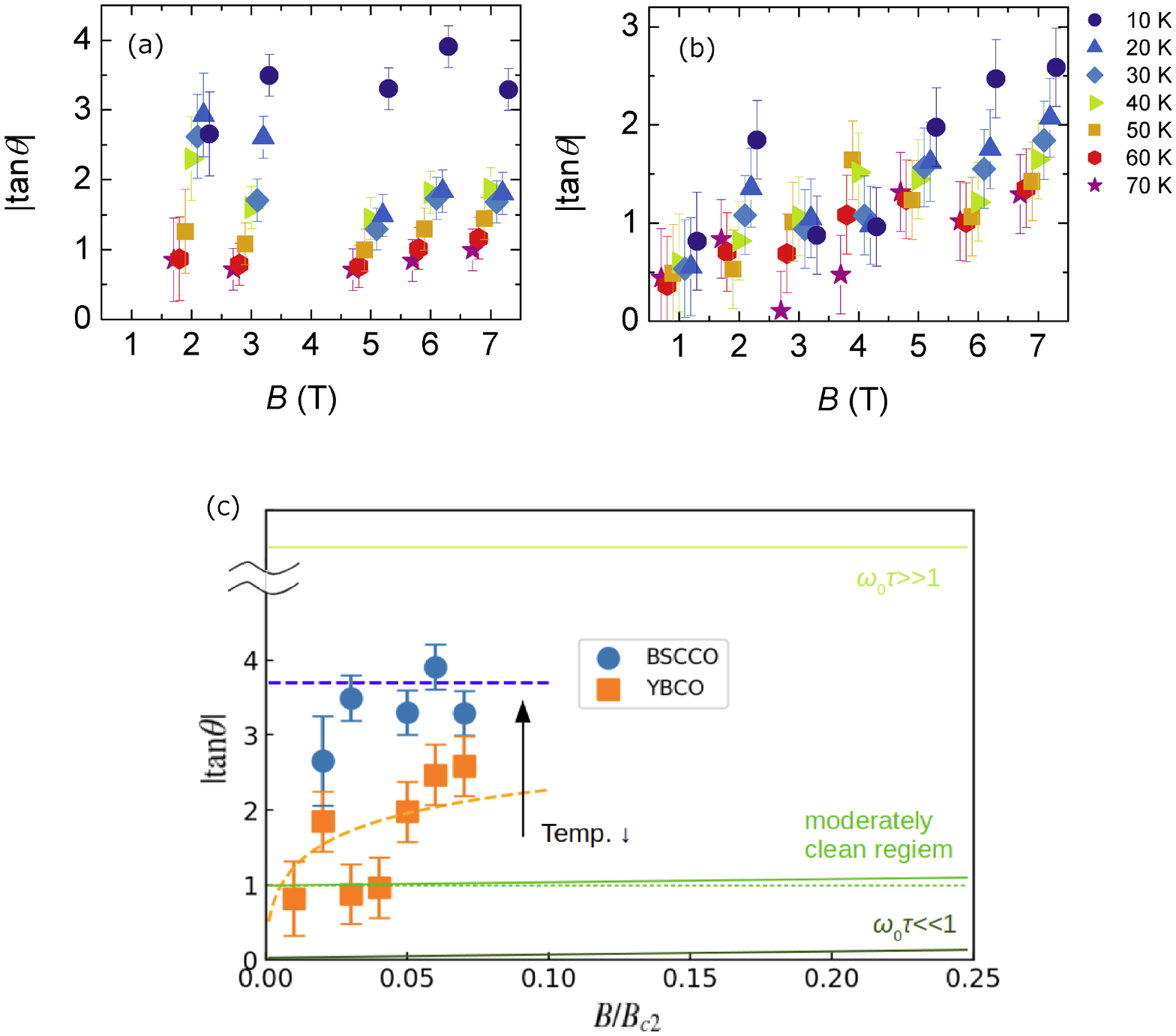}} 
	\caption{Tangent of the flux-flow Hall angle of (a) BSCCO and (b) YBCO as a function of the magnetic field from 10 K to 70 K.
The magnitude of the tangent of Hall angle increases with decreasing temperature, and it reaches an order of unity both for BSCCO and YBCO.
The tangent of Hall angle did not show remarkable dependence on the magnetic field for BSCCO, whereas it showed magnetic field dependence even at the lowest temperatures for YBCO.
(c) Summary of the experimental results at 10 K and theoretical expectations of FFHE.
Here, we assumed that $B_{c2}\sim100$ T both for BSCCO and YBCO.
The dashed lines are visual guides.
}
	\label{tan}
\end{figure}
Fig.~\ref{tan}(a) shows the tangent of the Hall angle of BSCCO obtained from Eq.~(\ref{method_tan}). 
The Hall angle increases with decreasing temperature and its tangent becomes approximately 3 at 10 K.
This is remarkable in two senses. 
First, this number is larger by one order of magnitude than those obtained previously from $\eta_{eff}$ by using the ordinary cavity perturbation technique~\cite{Hanaguri1999}. 
Second, the Hall angle does not considerably depend on the magnetic field, which is expected for a clean core.
This suggests that the vortex core of BSCCO is at least cleaner than expected previously based on $\eta_{eff}$.
Fig.~\ref{tan}(b) shows the tangent of the Hall angle of YBCO.
The tangent of the Hall angle of YBCO reaches an order of unity at 10 K, which is also one order of magnitude larger than those obtained by $\eta_{eff}$~\cite{Tsuchiya2001}.
Another interesting feature of the data in YBCO is that the Hall angle shows remarkable magnetic field dependence even at the lowest temperatures, which increases with the magnetic field, whereas the Hall angle does not depend on the magnetic field in BSCCO (Fig.~\ref{tan}(c)).

\begin{figure}[htbp]
\centering
 	\includegraphics[bb=50 150 850 450, keepaspectratio,width=85mm]{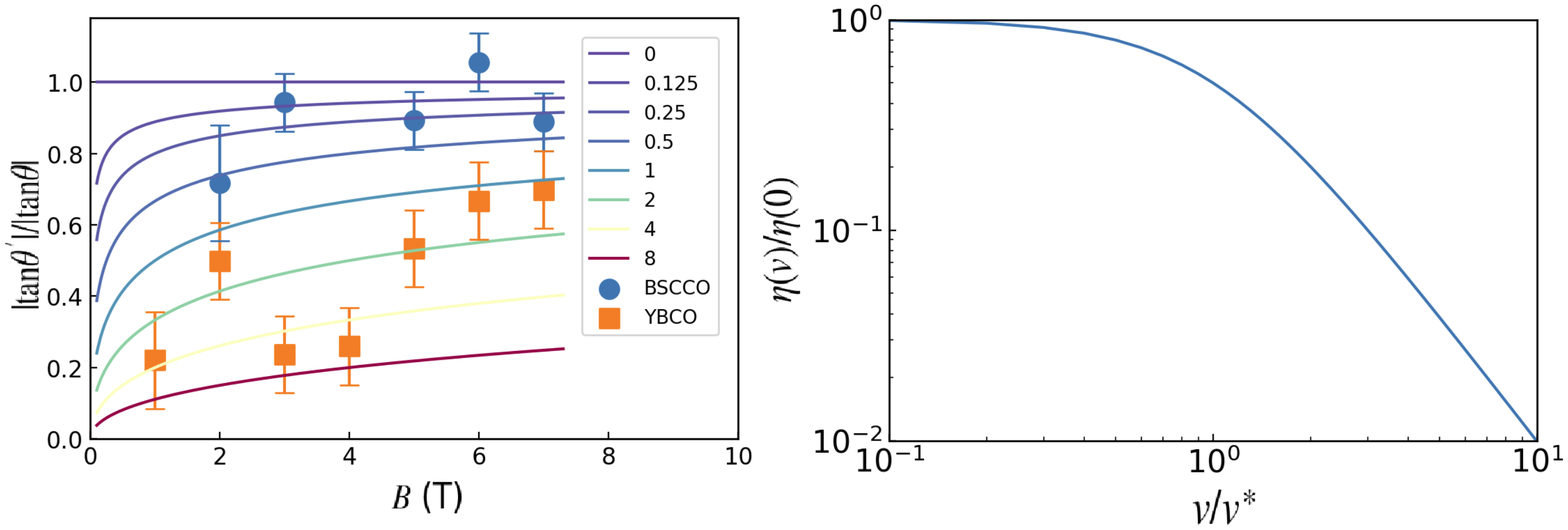}
 	\caption{(a) Magnetic field dependence of the absolute value of the Hall angle for vaious $K=\omega_p/\omega$. Here, we assume that the pinning constant $k_p$ depends on magnetic field as $B^{-1/2}$ and $|\tan\theta|$ equals to 3.7 both for BSCCO and YBCO.When the effect of pinning can be ignored ($K=0$), the Hall angle is constant. On the other hand, when the effect of pinning is present ($K\ne 0$), the Hall angle increases with increasing magnetic field.(b) The velocity dependence of viscous drag coefficient. In the case that $v$ is three times larger than $v^{\ast}$, the viscous darag coefficient is more than one order of magnitude smaller from that in the low velocity limit.}
 	\label{Fa_all}
\end{figure}

First, we discuss the latter feature.
We believe that the difference in the magnetic field dependence of the low-temperature Hall angle between BSCCO and YBCO is because of the remarkable difference in the equilibrium $B$-$T$ phase diagram.
In BSCCO, where the anisotropy is extremely strong, the vortices are in the liquid state for the most part of temperature and magnetic field~\cite{Lee1993,Zeldov1995}, whereas in YBCO, they are mostly in the solid phase~\cite{Shibata2002}.
Thus, the observed difference is likely due to the difference in the influence of the pinning. 
In fact, solving Eq.~(\ref{eqm}) yields
\begin{equation}\label{|tan|}
|\tan\theta^{\prime}|=|\tan\theta|\times\left|1+i\left(\omega_p/\omega\right)\right|^{-1},
\end{equation}
where $\tan\theta^{\prime}$ indicates that it is the observed value under the presence of pinning, whereas $\tan\theta$ is for the one without pinning. 
Generally, the pinning becomes less effective with increasing magnetic field. 
We represent the strength of the pinning by a parameter $K\equiv B^{\frac{1}{2}}\omega_{p}/\omega$, where $\omega_p=\kappa_p/\eta$ is the crossover frequency from the reactive motion to the dissipative motion and $\omega$ is the angular frequency, respectively.  For example, high-$T_c$ cuprate superconductors in vortex solid state have $\omega_p\sim 50$ GHz \cite{Tsuchiya2001}, that is, $K>1$. 
It should be noted that $K$ is zero in the vortex liquid state, which means that in BSCCO $K = 0$ in most part of the phase diagram.
The behavior of $|\tan\theta^{\prime}|$ expected by Eq.~(\ref{|tan|}) is shown as Fig.~\ref{Fa_all}(a) with an assumption that $\kappa_p$ depends on magnetic field as $B^{-1/2}$~\cite{Campbell1971}, which means that $K$ is field independent, and $|\tan\theta|=3.7$.
The equation reproduces the tendency that with increasing magnetic field, $|\tan\theta^{\prime}/\tan\theta|$ increases.  The magnitude of the parameter used is also reasonable; for the data of YBCO at 10 K, 
 the experimental results is in agreement with the curve with $K = 2$. 
Note that the effect of pinning on the Hall angle in the mixed state has also been discussed in terms of anomalous sign reversal of Hall resistivity~\cite{Kopnin1999}.
Other possibility to understand the difference between BSCCO and YBCO is the difference of vortex feature (Abrikosov vs pan cake).
If we consider stack of pan-cake vortices, it might be possible that pan cakes in each layer oscillates incoherently, which can make the Hall signal smaller than is expected for a single Abrikosov vortex. 
However, it does not lead to the observation that the Hall angle in BSCCO is larger than that in YBCO. 

Next, we discuss the most remarkable but puzzling feature of the present experiment; the large Hall angle. 
The observed large Hall angle is considered to be a common feature of the flux flow of high-$T_c$ cuprate superconductors since both BSCCO and YBCO show a similar number, instead of the fairly different features of vortices themselves.
First, the temperature dependences are similar to those observed in the effective viscous drag coefficient study~\cite{Tsuchiya2001}; it increased with decreasing temperature.
This has been understood by the increase in the superconducting gap at low temperatures, since $\tan\theta=\Delta^2\tau/\hbar E_F$.
The puzzling feature is its magnitude. 
As discussed above, the field independent behavior observed in BSCCO represents the more essential feature of  FFHE of high-$T_c$ cuprates without pinning.
This suggests that the vortex core of the cuprate superconductors is cleaner ($r=\omega_0\tau\sim$1) than expected in previous experiments~\cite{Tsuchiya2001,Shibata2003,Hanaguri1999,Maeda2007a,Maeda2007},
but it is rather consistent with the early expected clean core for superconductors with the large $\Delta$ and the small $E_F$.
Thus, the most serious problem to be solved is the disagreement of the Hall angle obtained in the two different techniques.

There have been long-term extensive discussions on the forces acting on the vortex and the resultant equation of motion. However, the exact descriptions of the forces acting on the vortex in superconductors have not been obtained~\cite{Bardeen1965,Nozieres1966,Gorkov1976,Hu1972,Dorsey1992,Ao1993,Sonin1997,Chen1998,Kopnin2002,Narayan2003,Kato2016a}. 
In terms of this, we can consider the differences as follows. 
The effective viscous drag coefficient is defined by Eq.~(3). 
It is related to the viscous drag coefficient of the longitudinal direction and the ratio of the viscous drag coefficient in the transverse direction to the longitudinal direction. 
On the other hand, flux-flow Hall angle is defined by the Eq.~(2).
 It is related to the ratio of the viscous drag coefficients of the transverse direction to the longitudinal direction alone.
Therefore, if there is anything which was considered not to be important nor is considered at all, it might be possible to understand the difference.
This can be expressed as the following simple model.
For some reason, suppose that the measured viscous drag coefficients in the longitudinal $\eta^{*}$ and transverse directions $\alpha^{*}$ comparably become smaller than those directly tied to the QP state that have been considered so far, that is, $\eta$ and $\alpha$. 
This assumption yields
$
\eta^{*} =C\eta
$
and 
$
\alpha_{H}^{*} =C\alpha_{H},
$
where $C$ is a constant less than unity.
We find the measured flux-flow Hall angle as 
\begin{equation}\label{eq:conc11}
\tan\theta^{*}=\frac{\alpha_{H}^{*}}{\eta^{*}}=\frac{\alpha_{H}}{\eta}=\tan\theta.
\end{equation}
On the other hand, we find the effective viscous drag coefficient as 
\begin{equation}\label{eq:conc12}
\eta^{*}_{eff}=\eta^{*}+\frac{{\alpha_{H}^{*}}^2}{\eta^{*}}=C\left(\eta+\frac{\alpha_{H}^2}{\eta}\right) = C\eta_{eff}< \eta_{eff}.
\end{equation}
As above, the factor $C$ disappears in the measured flux-flow Hall angle $\tan\theta^{\ast}$; however, it remains in the measured effective viscous drag coefficient $\eta^{\ast}_{eff}$.
Thus, $\omega_0\tau$ estimated from the effective viscous drag coefficient is smaller than that evaluated from the FFHE measurements, and this result is consistent with the experimental results.

Then, the central issue to be solved is the origin of $C$.
One possibility for the reduction of the viscous drag coefficient is  the nonlinearity, namely, velocity dependence of the viscosity drag coefficient.
Larkin and Ovchinnikov showed that a high velocity results in the decrease in $\eta$ as $\eta(v)=\eta(0)/[1+(v/v^{\ast})^2]$, where a characteristic velocity $v^{\ast}$ is proportional to $(1-T/T_c)^{\frac{1}{4}}\tau_{in}^{-\frac{1}{2}}$ with an inelastic electron-electron scattering time $\tau_{in}$~\cite{LarkinA.I.;Ovchinnikov1976}.
Thus, for example, if $v\geq 3v^{\ast}$, the viscous drag coefficient is different by more than one order of magnitude from that in the low velocity limit, which is tied to the QPs state in the vortex core~(Fig.~\ref{Fa_all}(b)).  
In BSCCO and YBCO, $v^{\ast}$ at 70 K was estimated to be $\sim10^4$ cm/s by the $I$-$V$ characteristic study~\cite{Xiao1998,Doettinger1994}.
In contrast, our microwave electric field was typically $\sim 10^{-2}$ V/cm, leading to the vortex velocity $v=E/B\sim 10^2$ cm/s.
Although this number is seemingly less than $v^{\ast}$, we cannot exclude this possibility by considering that
 $v^{\ast}$ decrease as temperature decrease~\cite{Leo2011}.
In the paper, the quasiparticle relaxation time of a superconductor for various cleanness is calculated using Eliashberg formalism. 
They showed the temperature dependence of the critical velocity $v^{\ast}$ as a function of temperature, for various QP relation time $\tau_{\epsilon}$ which is expressed as $[(1/\tau_s) + (1/\tau_r)]^{-1}$, where $\tau_s$ and $\tau_r$ are scattering time of the QP and the recommbination time of the cooper pair, respectively. 
 For $\tau_{\epsilon}\approx\tau_r$,  $v^{\ast}$ decreases more than two orders of magnitude smaller from 0.8 $T_c$ to 0.1 $T_c$.  
 In cuprates, it is well established that the quasiparticle scattering time increases sharply with decreasing temperature~\cite{Bonn1992}, whereas it is reasonable to assume that $\tau_r$ is almost equal to the one for andreev reflection at the vortex core boundary.  
 Thus, it is likely that $\tau_{\epsilon}\approx\tau_r$ in our experimental situation for most of the temperature region investigated.
Therefore, $v^{\ast}$ at low temperature may be larger than $v$ and we cannot exclude the possibility that the viscous drag coefficient decreases at low temperatures due to the velocity dependence.

Another possibility is the existence of a {\it hidden} mechanism of dissipation; some dissipation mechanisms  which cannot be represented in the simple equation of motion, but they do contribute to the vortex motion.
Possible candidates might be the dissipations at the core boundary, which has not been understood well even now~\cite{Hofmann1998,Eschrig1999,Volovik1993,Hayashi1998}.
Also, it might be related to the recently proposed new dissipation mechanism in flux flow regime~\cite{smith2020}.
Moreover, experimental results suggesting the possibility of the additional dissipation have been obtained from measurements not only on cuprate superconductors but also on iron-based superconductors~\cite{Okada2012}.
In short, the discrepancy of two measurements can be due to the velocity dependence of viscous drag coefficient and/or the additional dissipation, which has not been considered so far. 
Regardless, the unexpectedly large Hall angle of the vortex motion in cuprates revealed in our FFHE study poses a serious question on the fundamental understanding of the motion of the quantized vortex in superconductors.
Therefore, it is extremely important how large the Hall angle measured in this technique is for other superconductors, not only for which a clean core feature is expected, such as FeSe~\cite{Hanaguri2019}, but even for many conventional superconductors to perform a quantitative evaluation and identify its origins.

In conclusion, we investigated the nature of the QP state in the vortex core by means of the FFHE measurements at 15.8 GHz. 
We measured FFHE in cuprate superconductors, BSCCO and YBCO single crystals.
Both materials showed a large tangent of the Hall angle at low temperatures, which was larger by an order of magnitude than those obtained in the effective viscous drag coefficient measurement. 
We discussed the origin of the discrepancy both  in terms of the possible nonlinearity of the viscous drag force and the possible hidden dissipation mechanisms.
The unexpectedly large Hall angle of the vortex motion in cuprates revealed in our FFHE study poses a serious question on the fundamental understanding of the motion of the quantized vortex, and it deserves further investigation.

\begin{acknowledgments}
We thank Professor Y. Kato for enlightening discussions.
T.N. acknowledges support from a Grant-in-Aid for Scientific 
Research (KAKENHI) (project no. 20K03867) and the Advanced 
Instruments Center of Kyushu Sangyo University.
\end{acknowledgments}


\begin{thebibliography}{67}
\expandafter\ifx\csname natexlab\endcsname\relax\def\natexlab#1{#1}\fi
\expandafter\ifx\csname bibnamefont\endcsname\relax
  \def\bibnamefont#1{#1}\fi
\expandafter\ifx\csname bibfnamefont\endcsname\relax
  \def\bibfnamefont#1{#1}\fi
\expandafter\ifx\csname citenamefont\endcsname\relax
  \def\citenamefont#1{#1}\fi
\expandafter\ifx\csname url\endcsname\relax
  \def\url#1{\texttt{#1}}\fi
\expandafter\ifx\csname urlprefix\endcsname\relax\def\urlprefix{URL }\fi
\providecommand{\bibinfo}[2]{#2}
\providecommand{\eprint}[2][]{\url{#2}}

\bibitem[{\citenamefont{Caroli et~al.}(1964)\citenamefont{Caroli, {De Gennes},
  and Matricon}}]{Caroli1964}
\bibinfo{author}{\bibfnamefont{C.}~\bibnamefont{Caroli}},
  \bibinfo{author}{\bibfnamefont{P.}~\bibnamefont{{De Gennes}}},
  \bibnamefont{and} \bibinfo{author}{\bibfnamefont{J.}~\bibnamefont{Matricon}},
  \bibinfo{journal}{Physics Letters} \textbf{\bibinfo{volume}{9}},
  \bibinfo{pages}{307} (\bibinfo{year}{1964}).

\bibitem[{\citenamefont{Maggio-Aprile et~al.}(1995)\citenamefont{Maggio-Aprile,
  Renner, Erb, Walker, and Fischer}}]{Maggio-Aprile1995}
\bibinfo{author}{\bibfnamefont{I.}~\bibnamefont{Maggio-Aprile}},
  \bibinfo{author}{\bibfnamefont{C.}~\bibnamefont{Renner}},
  \bibinfo{author}{\bibfnamefont{A.}~\bibnamefont{Erb}},
  \bibinfo{author}{\bibfnamefont{E.}~\bibnamefont{Walker}}, \bibnamefont{and}
  \bibinfo{author}{\bibfnamefont{{\O}.}~\bibnamefont{Fischer}},
  \bibinfo{journal}{Physical Review Letters} \textbf{\bibinfo{volume}{75}},
  \bibinfo{pages}{2754} (\bibinfo{year}{1995}).

\bibitem[{\citenamefont{Pan et~al.}(2000)\citenamefont{Pan, Hudson, Gupta, Ng,
  Eisaki, Uchida, and Davis}}]{Pan2000}
\bibinfo{author}{\bibfnamefont{S.~H.} \bibnamefont{Pan}},
  \bibinfo{author}{\bibfnamefont{E.~W.} \bibnamefont{Hudson}},
  \bibinfo{author}{\bibfnamefont{A.~K.} \bibnamefont{Gupta}},
  \bibinfo{author}{\bibfnamefont{K.-W.} \bibnamefont{Ng}},
  \bibinfo{author}{\bibfnamefont{H.}~\bibnamefont{Eisaki}},
  \bibinfo{author}{\bibfnamefont{S.}~\bibnamefont{Uchida}}, \bibnamefont{and}
  \bibinfo{author}{\bibfnamefont{J.~C.} \bibnamefont{Davis}},
  \bibinfo{journal}{Physical Review Letters} \textbf{\bibinfo{volume}{85}},
  \bibinfo{pages}{1536} (\bibinfo{year}{2000}).

\bibitem[{\citenamefont{Valles et~al.}(1991)\citenamefont{Valles, Dynes,
  Cucolo, Gurvitch, Schneemeyer, Garno, and Waszczak}}]{Valles1991}
\bibinfo{author}{\bibfnamefont{J.~M.} \bibnamefont{Valles}},
  \bibinfo{author}{\bibfnamefont{R.~C.} \bibnamefont{Dynes}},
  \bibinfo{author}{\bibfnamefont{A.~M.} \bibnamefont{Cucolo}},
  \bibinfo{author}{\bibfnamefont{M.}~\bibnamefont{Gurvitch}},
  \bibinfo{author}{\bibfnamefont{L.~F.} \bibnamefont{Schneemeyer}},
  \bibinfo{author}{\bibfnamefont{J.~P.} \bibnamefont{Garno}}, \bibnamefont{and}
  \bibinfo{author}{\bibfnamefont{J.~V.} \bibnamefont{Waszczak}},
  \bibinfo{journal}{Physical Review B} \textbf{\bibinfo{volume}{44}},
  \bibinfo{pages}{11986} (\bibinfo{year}{1991}).

\bibitem[{\citenamefont{Edwards et~al.}(1992)\citenamefont{Edwards, Markert,
  and de~Lozanne}}]{Edwards1992}
\bibinfo{author}{\bibfnamefont{H.~L.} \bibnamefont{Edwards}},
  \bibinfo{author}{\bibfnamefont{J.~T.} \bibnamefont{Markert}},
  \bibnamefont{and} \bibinfo{author}{\bibfnamefont{A.~L.}
  \bibnamefont{de~Lozanne}}, \bibinfo{journal}{Physical Review Letters}
  \textbf{\bibinfo{volume}{69}}, \bibinfo{pages}{2967} (\bibinfo{year}{1992}).

\bibitem[{\citenamefont{Miller et~al.}(1993)\citenamefont{Miller, McElfresh,
  and Reifenberger}}]{Miller1993}
\bibinfo{author}{\bibfnamefont{T.~G.} \bibnamefont{Miller}},
  \bibinfo{author}{\bibfnamefont{M.}~\bibnamefont{McElfresh}},
  \bibnamefont{and}
  \bibinfo{author}{\bibfnamefont{R.}~\bibnamefont{Reifenberger}},
  \bibinfo{journal}{Physical Review B} \textbf{\bibinfo{volume}{48}},
  \bibinfo{pages}{7499} (\bibinfo{year}{1993}).

\bibitem[{\citenamefont{Nantoh et~al.}(1994)\citenamefont{Nantoh, Hasegawa,
  Yamaguchi, Takagi, Ogino, Kitazawa, Kawasaki, Gong, and
  Koinuma}}]{Nantoh1994}
\bibinfo{author}{\bibfnamefont{M.}~\bibnamefont{Nantoh}},
  \bibinfo{author}{\bibfnamefont{T.}~\bibnamefont{Hasegawa}},
  \bibinfo{author}{\bibfnamefont{W.}~\bibnamefont{Yamaguchi}},
  \bibinfo{author}{\bibfnamefont{A.}~\bibnamefont{Takagi}},
  \bibinfo{author}{\bibfnamefont{M.}~\bibnamefont{Ogino}},
  \bibinfo{author}{\bibfnamefont{K.}~\bibnamefont{Kitazawa}},
  \bibinfo{author}{\bibfnamefont{M.}~\bibnamefont{Kawasaki}},
  \bibinfo{author}{\bibfnamefont{J.}~\bibnamefont{Gong}}, \bibnamefont{and}
  \bibinfo{author}{\bibfnamefont{H.}~\bibnamefont{Koinuma}},
  \bibinfo{journal}{Journal of Applied Physics} \textbf{\bibinfo{volume}{75}},
  \bibinfo{pages}{5227} (\bibinfo{year}{1994}).

\bibitem[{\citenamefont{Shibata
  et~al.}(2003{\natexlab{a}})\citenamefont{Shibata, Maki, Nishizaki, and
  Kobayashi}}]{Shibata2003a}
\bibinfo{author}{\bibfnamefont{K.}~\bibnamefont{Shibata}},
  \bibinfo{author}{\bibfnamefont{M.}~\bibnamefont{Maki}},
  \bibinfo{author}{\bibfnamefont{T.}~\bibnamefont{Nishizaki}},
  \bibnamefont{and}
  \bibinfo{author}{\bibfnamefont{N.}~\bibnamefont{Kobayashi}},
  \bibinfo{journal}{Physica C: Superconductivity}
  \textbf{\bibinfo{volume}{392-396}}, \bibinfo{pages}{323}
  (\bibinfo{year}{2003}{\natexlab{a}}).

\bibitem[{\citenamefont{Shibata et~al.}(2010)\citenamefont{Shibata, Nishizaki,
  Maki, and Kobayashi}}]{Shibata2010}
\bibinfo{author}{\bibfnamefont{K.}~\bibnamefont{Shibata}},
  \bibinfo{author}{\bibfnamefont{T.}~\bibnamefont{Nishizaki}},
  \bibinfo{author}{\bibfnamefont{M.}~\bibnamefont{Maki}}, \bibnamefont{and}
  \bibinfo{author}{\bibfnamefont{N.}~\bibnamefont{Kobayashi}},
  \bibinfo{journal}{Superconductor Science and Technology}
  \textbf{\bibinfo{volume}{23}}, \bibinfo{pages}{085004}
  (\bibinfo{year}{2010}).

\bibitem[{\citenamefont{Bru{\'{e}}r et~al.}(2016)\citenamefont{Bru{\'{e}}r,
  Maggio-Aprile, Jenkins, Risti{\'{c}}, Erb, Berthod, Fischer, and
  Renner}}]{Bruer2016}
\bibinfo{author}{\bibfnamefont{J.}~\bibnamefont{Bru{\'{e}}r}},
  \bibinfo{author}{\bibfnamefont{I.}~\bibnamefont{Maggio-Aprile}},
  \bibinfo{author}{\bibfnamefont{N.}~\bibnamefont{Jenkins}},
  \bibinfo{author}{\bibfnamefont{Z.}~\bibnamefont{Risti{\'{c}}}},
  \bibinfo{author}{\bibfnamefont{A.}~\bibnamefont{Erb}},
  \bibinfo{author}{\bibfnamefont{C.}~\bibnamefont{Berthod}},
  \bibinfo{author}{\bibfnamefont{{\O}.}~\bibnamefont{Fischer}},
  \bibnamefont{and} \bibinfo{author}{\bibfnamefont{C.}~\bibnamefont{Renner}},
  \bibinfo{journal}{Nature Communications} \textbf{\bibinfo{volume}{7}},
  \bibinfo{pages}{11139} (\bibinfo{year}{2016}).

\bibitem[{\citenamefont{Berthod et~al.}(2017)\citenamefont{Berthod,
  Maggio-Aprile, Bru{\'{e}}r, Erb, and Renner}}]{Berthod2017}
\bibinfo{author}{\bibfnamefont{C.}~\bibnamefont{Berthod}},
  \bibinfo{author}{\bibfnamefont{I.}~\bibnamefont{Maggio-Aprile}},
  \bibinfo{author}{\bibfnamefont{J.}~\bibnamefont{Bru{\'{e}}r}},
  \bibinfo{author}{\bibfnamefont{A.}~\bibnamefont{Erb}}, \bibnamefont{and}
  \bibinfo{author}{\bibfnamefont{C.}~\bibnamefont{Renner}},
  \bibinfo{journal}{Physical Review Letters} \textbf{\bibinfo{volume}{119}},
  \bibinfo{pages}{1} (\bibinfo{year}{2017}).

\bibitem[{\citenamefont{Renner et~al.}(1998)\citenamefont{Renner, Revaz,
  Kadowaki, Maggio-Aprile, and Fischer}}]{Renner1998}
\bibinfo{author}{\bibfnamefont{C.}~\bibnamefont{Renner}},
  \bibinfo{author}{\bibfnamefont{B.}~\bibnamefont{Revaz}},
  \bibinfo{author}{\bibfnamefont{K.}~\bibnamefont{Kadowaki}},
  \bibinfo{author}{\bibfnamefont{I.}~\bibnamefont{Maggio-Aprile}},
  \bibnamefont{and}
  \bibinfo{author}{\bibfnamefont{{\O}.}~\bibnamefont{Fischer}},
  \bibinfo{journal}{Physical Review Letters} \textbf{\bibinfo{volume}{80}},
  \bibinfo{pages}{3606} (\bibinfo{year}{1998}).

\bibitem[{\citenamefont{Hoogenboom et~al.}(2000)\citenamefont{Hoogenboom,
  Renner, Revaz, Maggio-Aprile, and Fischer}}]{Hoogenboom2000}
\bibinfo{author}{\bibfnamefont{B.~W.} \bibnamefont{Hoogenboom}},
  \bibinfo{author}{\bibfnamefont{C.}~\bibnamefont{Renner}},
  \bibinfo{author}{\bibfnamefont{B.}~\bibnamefont{Revaz}},
  \bibinfo{author}{\bibfnamefont{I.}~\bibnamefont{Maggio-Aprile}},
  \bibnamefont{and}
  \bibinfo{author}{\bibfnamefont{{\O}.}~\bibnamefont{Fischer}},
  \bibinfo{journal}{Physica C: Superconductivity}
  \textbf{\bibinfo{volume}{332}}, \bibinfo{pages}{440} (\bibinfo{year}{2000}).

\bibitem[{\citenamefont{Matsuba et~al.}(2003)\citenamefont{Matsuba, Sakata,
  Kosugi, Nishimori, and Nishida}}]{Matsuba2003}
\bibinfo{author}{\bibfnamefont{K.}~\bibnamefont{Matsuba}},
  \bibinfo{author}{\bibfnamefont{H.}~\bibnamefont{Sakata}},
  \bibinfo{author}{\bibfnamefont{N.}~\bibnamefont{Kosugi}},
  \bibinfo{author}{\bibfnamefont{H.}~\bibnamefont{Nishimori}},
  \bibnamefont{and} \bibinfo{author}{\bibfnamefont{N.}~\bibnamefont{Nishida}},
  \bibinfo{journal}{Journal of the Physical Society of Japan}
  \textbf{\bibinfo{volume}{72}}, \bibinfo{pages}{2153} (\bibinfo{year}{2003}).

\bibitem[{\citenamefont{Levy et~al.}(2005)\citenamefont{Levy, Kugler, Manuel,
  Fischer, and Li}}]{Levy2005}
\bibinfo{author}{\bibfnamefont{G.}~\bibnamefont{Levy}},
  \bibinfo{author}{\bibfnamefont{M.}~\bibnamefont{Kugler}},
  \bibinfo{author}{\bibfnamefont{A.~A.} \bibnamefont{Manuel}},
  \bibinfo{author}{\bibfnamefont{{\O}.}~\bibnamefont{Fischer}},
  \bibnamefont{and} \bibinfo{author}{\bibfnamefont{M.}~\bibnamefont{Li}},
  \bibinfo{journal}{Physical Review Letters} \textbf{\bibinfo{volume}{95}},
  \bibinfo{pages}{257005} (\bibinfo{year}{2005}).

\bibitem[{\citenamefont{Matsuba et~al.}(2007)\citenamefont{Matsuba, Yoshizawa,
  Mochizuki, Mochiku, Hirata, and Nishida}}]{Matsuba2007}
\bibinfo{author}{\bibfnamefont{K.}~\bibnamefont{Matsuba}},
  \bibinfo{author}{\bibfnamefont{S.}~\bibnamefont{Yoshizawa}},
  \bibinfo{author}{\bibfnamefont{Y.}~\bibnamefont{Mochizuki}},
  \bibinfo{author}{\bibfnamefont{T.}~\bibnamefont{Mochiku}},
  \bibinfo{author}{\bibfnamefont{K.}~\bibnamefont{Hirata}}, \bibnamefont{and}
  \bibinfo{author}{\bibfnamefont{N.}~\bibnamefont{Nishida}},
  \bibinfo{journal}{Journal of the Physical Society of Japan}
  \textbf{\bibinfo{volume}{76}}, \bibinfo{pages}{063704}
  (\bibinfo{year}{2007}).

\bibitem[{\citenamefont{Yoshizawa et~al.}(2013)\citenamefont{Yoshizawa, Koseki,
  Matsuba, Mochiku, Hirata, and Nishida}}]{Yoshizawa2013}
\bibinfo{author}{\bibfnamefont{S.}~\bibnamefont{Yoshizawa}},
  \bibinfo{author}{\bibfnamefont{T.}~\bibnamefont{Koseki}},
  \bibinfo{author}{\bibfnamefont{K.}~\bibnamefont{Matsuba}},
  \bibinfo{author}{\bibfnamefont{T.}~\bibnamefont{Mochiku}},
  \bibinfo{author}{\bibfnamefont{K.}~\bibnamefont{Hirata}}, \bibnamefont{and}
  \bibinfo{author}{\bibfnamefont{N.}~\bibnamefont{Nishida}},
  \bibinfo{journal}{Journal of the Physical Society of Japan}
  \textbf{\bibinfo{volume}{82}}, \bibinfo{pages}{083706}
  (\bibinfo{year}{2013}).

\bibitem[{\citenamefont{Bardeen and Stephen}(1965)}]{Bardeen1965}
\bibinfo{author}{\bibfnamefont{J.}~\bibnamefont{Bardeen}} \bibnamefont{and}
  \bibinfo{author}{\bibfnamefont{M.~J.} \bibnamefont{Stephen}},
  \bibinfo{journal}{Physical Review} \textbf{\bibinfo{volume}{140}},
  \bibinfo{pages}{A1197} (\bibinfo{year}{1965}).

\bibitem[{\citenamefont{Nozi{\`{e}}res and Vinen}(1966)}]{Nozieres1966}
\bibinfo{author}{\bibfnamefont{P.}~\bibnamefont{Nozi{\`{e}}res}}
  \bibnamefont{and} \bibinfo{author}{\bibfnamefont{W.~F.} \bibnamefont{Vinen}},
  \bibinfo{journal}{The Philosophical Magazine: A Journal of Theoretical
  Experimental and Applied Physics} \textbf{\bibinfo{volume}{14}},
  \bibinfo{pages}{667} (\bibinfo{year}{1966}).

\bibitem[{\citenamefont{Blatter et~al.}(1994)\citenamefont{Blatter, Feigel'man,
  Geshkenbein, Larkin, and Vinokur}}]{Blatter1994}
\bibinfo{author}{\bibfnamefont{G.}~\bibnamefont{Blatter}},
  \bibinfo{author}{\bibfnamefont{M.~V.} \bibnamefont{Feigel'man}},
  \bibinfo{author}{\bibfnamefont{V.~B.} \bibnamefont{Geshkenbein}},
  \bibinfo{author}{\bibfnamefont{A.~I.} \bibnamefont{Larkin}},
  \bibnamefont{and} \bibinfo{author}{\bibfnamefont{V.~M.}
  \bibnamefont{Vinokur}}, \bibinfo{journal}{Reviews of Modern Physics}
  \textbf{\bibinfo{volume}{66}}, \bibinfo{pages}{1125} (\bibinfo{year}{1994}).

\bibitem[{\citenamefont{Kopnin and Kravtsov}(1976{\natexlab{a}})}]{Kopnin1976}
\bibinfo{author}{\bibfnamefont{N.~B.} \bibnamefont{Kopnin}} \bibnamefont{and}
  \bibinfo{author}{\bibfnamefont{V.}~\bibnamefont{Kravtsov}},
  \bibinfo{journal}{Soviet Journal of Experimental and Theoretical Physics}
  \textbf{\bibinfo{volume}{44}}, \bibinfo{pages}{861}
  (\bibinfo{year}{1976}{\natexlab{a}}).

\bibitem[{\citenamefont{Golosovsky et~al.}(1994)\citenamefont{Golosovsky,
  Tsindlekht, Chayet, and Davidov}}]{Golosovsky1994}
\bibinfo{author}{\bibfnamefont{M.}~\bibnamefont{Golosovsky}},
  \bibinfo{author}{\bibfnamefont{M.}~\bibnamefont{Tsindlekht}},
  \bibinfo{author}{\bibfnamefont{H.}~\bibnamefont{Chayet}}, \bibnamefont{and}
  \bibinfo{author}{\bibfnamefont{D.}~\bibnamefont{Davidov}},
  \bibinfo{journal}{Physical Review B} \textbf{\bibinfo{volume}{50}},
  \bibinfo{pages}{470} (\bibinfo{year}{1994}).

\bibitem[{\citenamefont{Klein et~al.}(1993)\citenamefont{Klein, Donovan,
  Dressel, and Gr{\"{u}}ner}}]{Klein1993}
\bibinfo{author}{\bibfnamefont{O.}~\bibnamefont{Klein}},
  \bibinfo{author}{\bibfnamefont{S.}~\bibnamefont{Donovan}},
  \bibinfo{author}{\bibfnamefont{M.}~\bibnamefont{Dressel}}, \bibnamefont{and}
  \bibinfo{author}{\bibfnamefont{G.}~\bibnamefont{Gr{\"{u}}ner}},
  \bibinfo{journal}{International Journal of Infrared and Millimeter Waves}
  \textbf{\bibinfo{volume}{14}}, \bibinfo{pages}{2423} (\bibinfo{year}{1993}).

\bibitem[{\citenamefont{Donovan et~al.}(1993)\citenamefont{Donovan, Klein,
  Dressel, Holczer, and Gr{\"{u}}ner}}]{Donovan1993}
\bibinfo{author}{\bibfnamefont{S.}~\bibnamefont{Donovan}},
  \bibinfo{author}{\bibfnamefont{O.}~\bibnamefont{Klein}},
  \bibinfo{author}{\bibfnamefont{M.}~\bibnamefont{Dressel}},
  \bibinfo{author}{\bibfnamefont{K.}~\bibnamefont{Holczer}}, \bibnamefont{and}
  \bibinfo{author}{\bibfnamefont{G.}~\bibnamefont{Gr{\"{u}}ner}},
  \bibinfo{journal}{International Journal of Infrared and Millimeter Waves}
  \textbf{\bibinfo{volume}{14}}, \bibinfo{pages}{2459} (\bibinfo{year}{1993}).

\bibitem[{\citenamefont{Dressel et~al.}(1993)\citenamefont{Dressel, Klein,
  Donovan, and Gr{\"{u}}ner}}]{Dressel1993}
\bibinfo{author}{\bibfnamefont{M.}~\bibnamefont{Dressel}},
  \bibinfo{author}{\bibfnamefont{O.}~\bibnamefont{Klein}},
  \bibinfo{author}{\bibfnamefont{S.}~\bibnamefont{Donovan}}, \bibnamefont{and}
  \bibinfo{author}{\bibfnamefont{G.}~\bibnamefont{Gr{\"{u}}ner}},
  \bibinfo{journal}{International Journal of Infrared and Millimeter Waves}
  \textbf{\bibinfo{volume}{14}}, \bibinfo{pages}{2489} (\bibinfo{year}{1993}).

\bibitem[{\citenamefont{Tsuchiya et~al.}(2001)\citenamefont{Tsuchiya, Iwaya,
  Kinoshita, Hanaguri, Kitano, Maeda, Shibata, Nishizaki, and
  Kobayashi}}]{Tsuchiya2001}
\bibinfo{author}{\bibfnamefont{Y.}~\bibnamefont{Tsuchiya}},
  \bibinfo{author}{\bibfnamefont{K.}~\bibnamefont{Iwaya}},
  \bibinfo{author}{\bibfnamefont{K.}~\bibnamefont{Kinoshita}},
  \bibinfo{author}{\bibfnamefont{T.}~\bibnamefont{Hanaguri}},
  \bibinfo{author}{\bibfnamefont{H.}~\bibnamefont{Kitano}},
  \bibinfo{author}{\bibfnamefont{A.}~\bibnamefont{Maeda}},
  \bibinfo{author}{\bibfnamefont{K.}~\bibnamefont{Shibata}},
  \bibinfo{author}{\bibfnamefont{T.}~\bibnamefont{Nishizaki}},
  \bibnamefont{and}
  \bibinfo{author}{\bibfnamefont{N.}~\bibnamefont{Kobayashi}},
  \bibinfo{journal}{Physical Review B} \textbf{\bibinfo{volume}{63}},
  \bibinfo{pages}{184517} (\bibinfo{year}{2001}).

\bibitem[{\citenamefont{Shibata
  et~al.}(2003{\natexlab{b}})\citenamefont{Shibata, Matsumoto, Izawa, Matsuda,
  Lee, and Tajima}}]{Shibata2003}
\bibinfo{author}{\bibfnamefont{A.}~\bibnamefont{Shibata}},
  \bibinfo{author}{\bibfnamefont{M.}~\bibnamefont{Matsumoto}},
  \bibinfo{author}{\bibfnamefont{K.}~\bibnamefont{Izawa}},
  \bibinfo{author}{\bibfnamefont{Y.}~\bibnamefont{Matsuda}},
  \bibinfo{author}{\bibfnamefont{S.}~\bibnamefont{Lee}}, \bibnamefont{and}
  \bibinfo{author}{\bibfnamefont{S.}~\bibnamefont{Tajima}},
  \bibinfo{journal}{Physical Review B} \textbf{\bibinfo{volume}{68}},
  \bibinfo{pages}{060501} (\bibinfo{year}{2003}{\natexlab{b}}).

\bibitem[{\citenamefont{Hanaguri et~al.}(1999)\citenamefont{Hanaguri, Tsuboi,
  Tsuchiya, Sasaki, and Maeda}}]{Hanaguri1999}
\bibinfo{author}{\bibfnamefont{T.}~\bibnamefont{Hanaguri}},
  \bibinfo{author}{\bibfnamefont{T.}~\bibnamefont{Tsuboi}},
  \bibinfo{author}{\bibfnamefont{Y.}~\bibnamefont{Tsuchiya}},
  \bibinfo{author}{\bibfnamefont{K.}~\bibnamefont{Sasaki}}, \bibnamefont{and}
  \bibinfo{author}{\bibfnamefont{A.}~\bibnamefont{Maeda}},
  \bibinfo{journal}{Physical Review Letters} \textbf{\bibinfo{volume}{82}},
  \bibinfo{pages}{1273} (\bibinfo{year}{1999}).

\bibitem[{\citenamefont{Maeda et~al.}(2007{\natexlab{a}})\citenamefont{Maeda,
  Umetsu, and Kitano}}]{Maeda2007a}
\bibinfo{author}{\bibfnamefont{A.}~\bibnamefont{Maeda}},
  \bibinfo{author}{\bibfnamefont{T.}~\bibnamefont{Umetsu}}, \bibnamefont{and}
  \bibinfo{author}{\bibfnamefont{H.}~\bibnamefont{Kitano}},
  \bibinfo{journal}{Physica C: Superconductivity}
  \textbf{\bibinfo{volume}{460-462}}, \bibinfo{pages}{1202}
  (\bibinfo{year}{2007}{\natexlab{a}}).

\bibitem[{\citenamefont{Maeda et~al.}(2007{\natexlab{b}})\citenamefont{Maeda,
  Kitano, Kinoshita, Nishizaki, Shibata, and Kobayashi}}]{Maeda2007}
\bibinfo{author}{\bibfnamefont{A.}~\bibnamefont{Maeda}},
  \bibinfo{author}{\bibfnamefont{H.}~\bibnamefont{Kitano}},
  \bibinfo{author}{\bibfnamefont{K.}~\bibnamefont{Kinoshita}},
  \bibinfo{author}{\bibfnamefont{T.}~\bibnamefont{Nishizaki}},
  \bibinfo{author}{\bibfnamefont{K.}~\bibnamefont{Shibata}}, \bibnamefont{and}
  \bibinfo{author}{\bibfnamefont{N.}~\bibnamefont{Kobayashi}},
  \bibinfo{journal}{Journal of the Physical Society of Japan}
  \textbf{\bibinfo{volume}{76}}, \bibinfo{pages}{094708}
  (\bibinfo{year}{2007}{\natexlab{b}}).

\bibitem[{\citenamefont{Hofmann and K{\"{u}}mmel}(1998)}]{Hofmann1998}
\bibinfo{author}{\bibfnamefont{S.}~\bibnamefont{Hofmann}} \bibnamefont{and}
  \bibinfo{author}{\bibfnamefont{R.}~\bibnamefont{K{\"{u}}mmel}},
  \bibinfo{journal}{Physical Review B} \textbf{\bibinfo{volume}{57}},
  \bibinfo{pages}{7904} (\bibinfo{year}{1998}).

\bibitem[{\citenamefont{Eschrig et~al.}(1999)\citenamefont{Eschrig, Sauls, and
  Rainer}}]{Eschrig1999}
\bibinfo{author}{\bibfnamefont{M.}~\bibnamefont{Eschrig}},
  \bibinfo{author}{\bibfnamefont{J.~A.} \bibnamefont{Sauls}}, \bibnamefont{and}
  \bibinfo{author}{\bibfnamefont{D.}~\bibnamefont{Rainer}},
  \bibinfo{journal}{Physical Review B} \textbf{\bibinfo{volume}{60}},
  \bibinfo{pages}{10447} (\bibinfo{year}{1999}).

\bibitem[{\citenamefont{Volovik}(1993)}]{Volovik1993}
\bibinfo{author}{\bibfnamefont{G.~E.} \bibnamefont{Volovik}},
  \bibinfo{journal}{JETP} \textbf{\bibinfo{volume}{77}}, \bibinfo{pages}{435}
  (\bibinfo{year}{1993}).

\bibitem[{\citenamefont{Hayashi}(1998)}]{Hayashi1998}
\bibinfo{author}{\bibfnamefont{M.}~\bibnamefont{Hayashi}},
  \bibinfo{journal}{Journal of the Physical Society of Japan}
  \textbf{\bibinfo{volume}{67}}, \bibinfo{pages}{3372} (\bibinfo{year}{1998}).

\bibitem[{\citenamefont{Smith et~al.}(2020)\citenamefont{Smith, Andreev,
  Feigel'man, and Spivak}}]{smith2020}
\bibinfo{author}{\bibfnamefont{M.}~\bibnamefont{Smith}},
  \bibinfo{author}{\bibfnamefont{A.~V.} \bibnamefont{Andreev}},
  \bibinfo{author}{\bibfnamefont{M.~V.} \bibnamefont{Feigel'man}},
  \bibnamefont{and} \bibinfo{author}{\bibfnamefont{B.~Z.}
  \bibnamefont{Spivak}}, \bibinfo{journal}{Physical Review B}
  \textbf{\bibinfo{volume}{102}}, \bibinfo{pages}{180507}
  (\bibinfo{year}{2020}).

\bibitem[{\citenamefont{Kopnin and Kravtsov}(1976{\natexlab{b}})}]{Kopnin1976a}
\bibinfo{author}{\bibfnamefont{N.~B.} \bibnamefont{Kopnin}} \bibnamefont{and}
  \bibinfo{author}{\bibfnamefont{V.}~\bibnamefont{Kravtsov}},
  \bibinfo{journal}{JETP Lett.} \textbf{\bibinfo{volume}{23}},
  \bibinfo{pages}{578} (\bibinfo{year}{1976}{\natexlab{b}}).

\bibitem[{\citenamefont{Kopnin and Volovik}(1997)}]{Kopnin1997}
\bibinfo{author}{\bibfnamefont{N.~B.} \bibnamefont{Kopnin}} \bibnamefont{and}
  \bibinfo{author}{\bibfnamefont{G.~E.} \bibnamefont{Volovik}},
  \bibinfo{journal}{Physical Review Letters} \textbf{\bibinfo{volume}{79}},
  \bibinfo{pages}{1377} (\bibinfo{year}{1997}).

\bibitem[{\citenamefont{Larkin and Ovchinnikov}(1995)}]{Larkin1995}
\bibinfo{author}{\bibfnamefont{A.~I.} \bibnamefont{Larkin}} \bibnamefont{and}
  \bibinfo{author}{\bibfnamefont{Y.~N.} \bibnamefont{Ovchinnikov}},
  \bibinfo{journal}{Physical Review B} \textbf{\bibinfo{volume}{51}},
  \bibinfo{pages}{5965} (\bibinfo{year}{1995}).

\bibitem[{\citenamefont{Kopnin and Lopatin}(1995)}]{Kopnin1995}
\bibinfo{author}{\bibfnamefont{N.~B.} \bibnamefont{Kopnin}} \bibnamefont{and}
  \bibinfo{author}{\bibfnamefont{A.~V.} \bibnamefont{Lopatin}},
  \bibinfo{journal}{Physical Review B} \textbf{\bibinfo{volume}{51}},
  \bibinfo{pages}{15291} (\bibinfo{year}{1995}).

\bibitem[{\citenamefont{Arahata and Kato}(2014)}]{Arahata2014}
\bibinfo{author}{\bibfnamefont{E.}~\bibnamefont{Arahata}} \bibnamefont{and}
  \bibinfo{author}{\bibfnamefont{Y.}~\bibnamefont{Kato}},
  \bibinfo{journal}{Journal of Low Temperature Physics}
  \textbf{\bibinfo{volume}{175}}, \bibinfo{pages}{346} (\bibinfo{year}{2014}).

\bibitem[{\citenamefont{Kohno et~al.}(2016)\citenamefont{Kohno, Ueki, and
  Kita}}]{Kohno2016}
\bibinfo{author}{\bibfnamefont{W.}~\bibnamefont{Kohno}},
  \bibinfo{author}{\bibfnamefont{H.}~\bibnamefont{Ueki}}, \bibnamefont{and}
  \bibinfo{author}{\bibfnamefont{T.}~\bibnamefont{Kita}},
  \bibinfo{journal}{Journal of the Physical Society of Japan}
  \textbf{\bibinfo{volume}{85}}, \bibinfo{pages}{083705}
  (\bibinfo{year}{2016}).

\bibitem[{\citenamefont{Hagen et~al.}(1990)\citenamefont{Hagen, Lobb, Greene,
  Forrester, and Kang}}]{Hagen1990}
\bibinfo{author}{\bibfnamefont{S.~J.} \bibnamefont{Hagen}},
  \bibinfo{author}{\bibfnamefont{C.~J.} \bibnamefont{Lobb}},
  \bibinfo{author}{\bibfnamefont{R.~L.} \bibnamefont{Greene}},
  \bibinfo{author}{\bibfnamefont{M.~G.} \bibnamefont{Forrester}},
  \bibnamefont{and} \bibinfo{author}{\bibfnamefont{J.~H.} \bibnamefont{Kang}},
  \bibinfo{journal}{Physical Review B} \textbf{\bibinfo{volume}{41}},
  \bibinfo{pages}{11630} (\bibinfo{year}{1990}).

\bibitem[{\citenamefont{Nagaoka et~al.}(1998)\citenamefont{Nagaoka, Matsuda,
  Obara, Sawa, Terashima, Chong, Takano, and Suzuki}}]{Nagaoka1998}
\bibinfo{author}{\bibfnamefont{T.}~\bibnamefont{Nagaoka}},
  \bibinfo{author}{\bibfnamefont{Y.}~\bibnamefont{Matsuda}},
  \bibinfo{author}{\bibfnamefont{H.}~\bibnamefont{Obara}},
  \bibinfo{author}{\bibfnamefont{A.}~\bibnamefont{Sawa}},
  \bibinfo{author}{\bibfnamefont{T.}~\bibnamefont{Terashima}},
  \bibinfo{author}{\bibfnamefont{I.}~\bibnamefont{Chong}},
  \bibinfo{author}{\bibfnamefont{M.}~\bibnamefont{Takano}}, \bibnamefont{and}
  \bibinfo{author}{\bibfnamefont{M.}~\bibnamefont{Suzuki}},
  \bibinfo{journal}{Physical Review Letters} \textbf{\bibinfo{volume}{80}},
  \bibinfo{pages}{3594} (\bibinfo{year}{1998}).

\bibitem[{\citenamefont{Gittleman and Rosenblum}(1968)}]{Gittleman1968}
\bibinfo{author}{\bibfnamefont{J.~I.} \bibnamefont{Gittleman}}
  \bibnamefont{and}
  \bibinfo{author}{\bibfnamefont{B.}~\bibnamefont{Rosenblum}},
  \bibinfo{journal}{Journal of Applied Physics} \textbf{\bibinfo{volume}{39}},
  \bibinfo{pages}{2617} (\bibinfo{year}{1968}).

\bibitem[{\citenamefont{Coffey and Clem}(1991)}]{Coffey1991}
\bibinfo{author}{\bibfnamefont{M.~W.} \bibnamefont{Coffey}} \bibnamefont{and}
  \bibinfo{author}{\bibfnamefont{J.~R.} \bibnamefont{Clem}},
  \bibinfo{journal}{Physical Review Letters} \textbf{\bibinfo{volume}{67}},
  \bibinfo{pages}{386} (\bibinfo{year}{1991}).

\bibitem[{\citenamefont{Ogawa et~al.}(2021)\citenamefont{Ogawa, Okada,
  Takahashi, Nabeshima, and Maeda}}]{Ogawa2021}
\bibinfo{author}{\bibfnamefont{R.}~\bibnamefont{Ogawa}},
  \bibinfo{author}{\bibfnamefont{T.}~\bibnamefont{Okada}},
  \bibinfo{author}{\bibfnamefont{H.}~\bibnamefont{Takahashi}},
  \bibinfo{author}{\bibfnamefont{F.}~\bibnamefont{Nabeshima}},
  \bibnamefont{and} \bibinfo{author}{\bibfnamefont{A.}~\bibnamefont{Maeda}},
  \bibinfo{journal}{Journal of Applied Physics} \textbf{\bibinfo{volume}{129}},
  \bibinfo{pages}{015102} (\bibinfo{year}{2021}).

\bibitem[{\citenamefont{Lee et~al.}(1993)\citenamefont{Lee, Zimmermann, Keller,
  Warden, Savi{\'{c}}, Schauwecker, Zech, Cubitt, Forgan, Kes
  et~al.}}]{Lee1993}
\bibinfo{author}{\bibfnamefont{S.~L.} \bibnamefont{Lee}},
  \bibinfo{author}{\bibfnamefont{P.}~\bibnamefont{Zimmermann}},
  \bibinfo{author}{\bibfnamefont{H.}~\bibnamefont{Keller}},
  \bibinfo{author}{\bibfnamefont{M.}~\bibnamefont{Warden}},
  \bibinfo{author}{\bibfnamefont{I.~M.} \bibnamefont{Savi{\'{c}}}},
  \bibinfo{author}{\bibfnamefont{R.}~\bibnamefont{Schauwecker}},
  \bibinfo{author}{\bibfnamefont{D.}~\bibnamefont{Zech}},
  \bibinfo{author}{\bibfnamefont{R.}~\bibnamefont{Cubitt}},
  \bibinfo{author}{\bibfnamefont{E.~M.} \bibnamefont{Forgan}},
  \bibinfo{author}{\bibfnamefont{P.~H.} \bibnamefont{Kes}},
  \bibnamefont{et~al.}, \bibinfo{journal}{Physical Review Letters}
  \textbf{\bibinfo{volume}{71}}, \bibinfo{pages}{3862} (\bibinfo{year}{1993}).

\bibitem[{\citenamefont{Zeldov et~al.}(1995)\citenamefont{Zeldov, Majer,
  Konczykowski, Geshkenbein, Vinokur, and Shtrikman}}]{Zeldov1995}
\bibinfo{author}{\bibfnamefont{E.}~\bibnamefont{Zeldov}},
  \bibinfo{author}{\bibfnamefont{D.}~\bibnamefont{Majer}},
  \bibinfo{author}{\bibfnamefont{M.}~\bibnamefont{Konczykowski}},
  \bibinfo{author}{\bibfnamefont{V.~B.} \bibnamefont{Geshkenbein}},
  \bibinfo{author}{\bibfnamefont{V.~M.} \bibnamefont{Vinokur}},
  \bibnamefont{and}
  \bibinfo{author}{\bibfnamefont{H.}~\bibnamefont{Shtrikman}},
  \bibinfo{journal}{Nature} \textbf{\bibinfo{volume}{375}},
  \bibinfo{pages}{373} (\bibinfo{year}{1995}).

\bibitem[{\citenamefont{Shibata et~al.}(2002)\citenamefont{Shibata, Nishizaki,
  Sasaki, and Kobayashi}}]{Shibata2002}
\bibinfo{author}{\bibfnamefont{K.}~\bibnamefont{Shibata}},
  \bibinfo{author}{\bibfnamefont{T.}~\bibnamefont{Nishizaki}},
  \bibinfo{author}{\bibfnamefont{T.}~\bibnamefont{Sasaki}}, \bibnamefont{and}
  \bibinfo{author}{\bibfnamefont{N.}~\bibnamefont{Kobayashi}},
  \bibinfo{journal}{Physical Review B} \textbf{\bibinfo{volume}{66}},
  \bibinfo{pages}{214518} (\bibinfo{year}{2002}).

\bibitem[{\citenamefont{Campbell}(1971)}]{Campbell1971}
\bibinfo{author}{\bibfnamefont{A.~M.} \bibnamefont{Campbell}},
  \bibinfo{journal}{Journal of Physics C: Solid State Physics}
  \textbf{\bibinfo{volume}{4}}, \bibinfo{pages}{3186} (\bibinfo{year}{1971}).

\bibitem[{\citenamefont{Kopnin and Vinokur}(1999)}]{Kopnin1999}
\bibinfo{author}{\bibfnamefont{N.~B.} \bibnamefont{Kopnin}} \bibnamefont{and}
  \bibinfo{author}{\bibfnamefont{V.~M.} \bibnamefont{Vinokur}},
  \bibinfo{journal}{Physical Review Letters} \textbf{\bibinfo{volume}{83}},
  \bibinfo{pages}{4864} (\bibinfo{year}{1999}).

\bibitem[{\citenamefont{Gor'kov and Kopnin}(1976)}]{Gorkov1976}
\bibinfo{author}{\bibfnamefont{L.~P.} \bibnamefont{Gor'kov}} \bibnamefont{and}
  \bibinfo{author}{\bibfnamefont{N.~B.} \bibnamefont{Kopnin}},
  \bibinfo{journal}{Soviet Physics Uspekhi} \textbf{\bibinfo{volume}{18}},
  \bibinfo{pages}{496} (\bibinfo{year}{1976}).

\bibitem[{\citenamefont{Hu and Thompson}(1972)}]{Hu1972}
\bibinfo{author}{\bibfnamefont{C.-R.} \bibnamefont{Hu}} \bibnamefont{and}
  \bibinfo{author}{\bibfnamefont{R.~S.} \bibnamefont{Thompson}},
  \bibinfo{journal}{Physical Review B} \textbf{\bibinfo{volume}{6}},
  \bibinfo{pages}{110} (\bibinfo{year}{1972}).

\bibitem[{\citenamefont{Dorsey}(1992)}]{Dorsey1992}
\bibinfo{author}{\bibfnamefont{A.~T.} \bibnamefont{Dorsey}},
  \bibinfo{journal}{Physical Review B} \textbf{\bibinfo{volume}{46}},
  \bibinfo{pages}{8376} (\bibinfo{year}{1992}).

\bibitem[{\citenamefont{Ao and Thouless}(1993)}]{Ao1993}
\bibinfo{author}{\bibfnamefont{P.}~\bibnamefont{Ao}} \bibnamefont{and}
  \bibinfo{author}{\bibfnamefont{D.~J.} \bibnamefont{Thouless}},
  \bibinfo{journal}{Physical Review Letters} \textbf{\bibinfo{volume}{70}},
  \bibinfo{pages}{2158} (\bibinfo{year}{1993}).

\bibitem[{\citenamefont{Sonin}(1997)}]{Sonin1997}
\bibinfo{author}{\bibfnamefont{E.~B.} \bibnamefont{Sonin}},
  \bibinfo{journal}{Physical Review B} \textbf{\bibinfo{volume}{55}},
  \bibinfo{pages}{485} (\bibinfo{year}{1997}).

\bibitem[{\citenamefont{Chen et~al.}(1998)\citenamefont{Chen, Moreno, Hernando,
  Sanchez, and Li}}]{Chen1998}
\bibinfo{author}{\bibfnamefont{D.-X.} \bibnamefont{Chen}},
  \bibinfo{author}{\bibfnamefont{J.~J.} \bibnamefont{Moreno}},
  \bibinfo{author}{\bibfnamefont{A.}~\bibnamefont{Hernando}},
  \bibinfo{author}{\bibfnamefont{A.}~\bibnamefont{Sanchez}}, \bibnamefont{and}
  \bibinfo{author}{\bibfnamefont{B.-Z.} \bibnamefont{Li}},
  \bibinfo{journal}{Physical Review B} \textbf{\bibinfo{volume}{57}},
  \bibinfo{pages}{5059} (\bibinfo{year}{1998}).

\bibitem[{\citenamefont{Kopnin}(2002)}]{Kopnin2002}
\bibinfo{author}{\bibfnamefont{N.~B.} \bibnamefont{Kopnin}},
  \bibinfo{journal}{Reports on Progress in Physics}
  \textbf{\bibinfo{volume}{65}}, \bibinfo{pages}{1633} (\bibinfo{year}{2002}).

\bibitem[{\citenamefont{Narayan}(2003)}]{Narayan2003}
\bibinfo{author}{\bibfnamefont{O.}~\bibnamefont{Narayan}},
  \bibinfo{journal}{Journal of Physics A: Mathematical and General}
  \textbf{\bibinfo{volume}{36}}, \bibinfo{pages}{L373} (\bibinfo{year}{2003}).

\bibitem[{\citenamefont{Kato and Chung}(2016)}]{Kato2016a}
\bibinfo{author}{\bibfnamefont{Y.}~\bibnamefont{Kato}} \bibnamefont{and}
  \bibinfo{author}{\bibfnamefont{C.~K.} \bibnamefont{Chung}},
  \bibinfo{journal}{Journal of the Physical Society of Japan}
  \textbf{\bibinfo{volume}{85}} (\bibinfo{year}{2016}).

\bibitem[{\citenamefont{Larkin and
  Ovchinnikov}(1976)}]{LarkinA.I.;Ovchinnikov1976}
\bibinfo{author}{\bibfnamefont{A.~I.} \bibnamefont{Larkin}} \bibnamefont{and}
  \bibinfo{author}{\bibfnamefont{Y.~N.} \bibnamefont{Ovchinnikov}},
  \bibinfo{journal}{JETP Lett.} \textbf{\bibinfo{volume}{23}},
  \bibinfo{pages}{187} (\bibinfo{year}{1976}).

\bibitem[{\citenamefont{Xiao et~al.}(1998)\citenamefont{Xiao, {Voss-de Haan},
  Jakob, and Adrian}}]{Xiao1998}
\bibinfo{author}{\bibfnamefont{Z.~L.} \bibnamefont{Xiao}},
  \bibinfo{author}{\bibfnamefont{P.}~\bibnamefont{{Voss-de Haan}}},
  \bibinfo{author}{\bibfnamefont{G.}~\bibnamefont{Jakob}}, \bibnamefont{and}
  \bibinfo{author}{\bibfnamefont{H.}~\bibnamefont{Adrian}},
  \bibinfo{journal}{Physical Review B} \textbf{\bibinfo{volume}{57}},
  \bibinfo{pages}{R736} (\bibinfo{year}{1998}).

\bibitem[{\citenamefont{Doettinger et~al.}(1994)\citenamefont{Doettinger,
  Huebener, Gerdemann, K{\"{u}}hle, Anders, Tr{\"{a}}uble, and
  Vill{\'{e}}gier}}]{Doettinger1994}
\bibinfo{author}{\bibfnamefont{S.~G.} \bibnamefont{Doettinger}},
  \bibinfo{author}{\bibfnamefont{R.~P.} \bibnamefont{Huebener}},
  \bibinfo{author}{\bibfnamefont{R.}~\bibnamefont{Gerdemann}},
  \bibinfo{author}{\bibfnamefont{A.}~\bibnamefont{K{\"{u}}hle}},
  \bibinfo{author}{\bibfnamefont{S.}~\bibnamefont{Anders}},
  \bibinfo{author}{\bibfnamefont{T.~G.} \bibnamefont{Tr{\"{a}}uble}},
  \bibnamefont{and} \bibinfo{author}{\bibfnamefont{J.~C.}
  \bibnamefont{Vill{\'{e}}gier}}, \bibinfo{journal}{Physical Review Letters}
  \textbf{\bibinfo{volume}{73}}, \bibinfo{pages}{1691} (\bibinfo{year}{1994}).

\bibitem[{\citenamefont{Leo et~al.}(2011)\citenamefont{Leo, Grimaldi, Citro,
  Nigro, Pace, and Huebener}}]{Leo2011}
\bibinfo{author}{\bibfnamefont{A.}~\bibnamefont{Leo}},
  \bibinfo{author}{\bibfnamefont{G.}~\bibnamefont{Grimaldi}},
  \bibinfo{author}{\bibfnamefont{R.}~\bibnamefont{Citro}},
  \bibinfo{author}{\bibfnamefont{A.}~\bibnamefont{Nigro}},
  \bibinfo{author}{\bibfnamefont{S.}~\bibnamefont{Pace}}, \bibnamefont{and}
  \bibinfo{author}{\bibfnamefont{R.~P.} \bibnamefont{Huebener}},
  \bibinfo{journal}{Physical Review B} \textbf{\bibinfo{volume}{84}},
  \bibinfo{pages}{014536} (\bibinfo{year}{2011}).

\bibitem[{\citenamefont{Bonn et~al.}(1992)\citenamefont{Bonn, Dosanjh, Liang,
  and Hardy}}]{Bonn1992}
\bibinfo{author}{\bibfnamefont{D.~A.} \bibnamefont{Bonn}},
  \bibinfo{author}{\bibfnamefont{P.}~\bibnamefont{Dosanjh}},
  \bibinfo{author}{\bibfnamefont{R.}~\bibnamefont{Liang}}, \bibnamefont{and}
  \bibinfo{author}{\bibfnamefont{W.~N.} \bibnamefont{Hardy}},
  \bibinfo{journal}{Physical Review Letters} \textbf{\bibinfo{volume}{68}},
  \bibinfo{pages}{2390} (\bibinfo{year}{1992}).

\bibitem[{\citenamefont{Okada et~al.}(2012)\citenamefont{Okada, Takahashi,
  Imai, Kitagawa, Matsubayashi, Uwatoko, and Maeda}}]{Okada2012}
\bibinfo{author}{\bibfnamefont{T.}~\bibnamefont{Okada}},
  \bibinfo{author}{\bibfnamefont{H.}~\bibnamefont{Takahashi}},
  \bibinfo{author}{\bibfnamefont{Y.}~\bibnamefont{Imai}},
  \bibinfo{author}{\bibfnamefont{K.}~\bibnamefont{Kitagawa}},
  \bibinfo{author}{\bibfnamefont{K.}~\bibnamefont{Matsubayashi}},
  \bibinfo{author}{\bibfnamefont{Y.}~\bibnamefont{Uwatoko}}, \bibnamefont{and}
  \bibinfo{author}{\bibfnamefont{A.}~\bibnamefont{Maeda}},
  \bibinfo{journal}{Physical Review B} \textbf{\bibinfo{volume}{86}},
  \bibinfo{pages}{064516} (\bibinfo{year}{2012}).

\bibitem[{\citenamefont{Hanaguri et~al.}(2019)\citenamefont{Hanaguri, Kasahara,
  B{\"{o}}ker, Eremin, Shibauchi, and Matsuda}}]{Hanaguri2019}
\bibinfo{author}{\bibfnamefont{T.}~\bibnamefont{Hanaguri}},
  \bibinfo{author}{\bibfnamefont{S.}~\bibnamefont{Kasahara}},
  \bibinfo{author}{\bibfnamefont{J.}~\bibnamefont{B{\"{o}}ker}},
  \bibinfo{author}{\bibfnamefont{I.}~\bibnamefont{Eremin}},
  \bibinfo{author}{\bibfnamefont{T.}~\bibnamefont{Shibauchi}},
  \bibnamefont{and} \bibinfo{author}{\bibfnamefont{Y.}~\bibnamefont{Matsuda}},
  \bibinfo{journal}{Physical Review Letters} \textbf{\bibinfo{volume}{122}},
  \bibinfo{pages}{077001} (\bibinfo{year}{2019}).

\end{thebibliography}
%

\end{document}